\documentclass[submission, Proceedings]{SciPost}

\hypersetup{
    colorlinks,
    linkcolor={red!50!black},
    citecolor={blue!50!black},
    urlcolor={blue!80!black}
}

\usepackage[bitstream-charter]{mathdesign}
\usepackage[compat=1.1.0]{tikz-feynman}
\newcommand{\be}{\begin{equation}}
\newcommand{\ee}{\end{equation}}
\newcommand{\bea}{\begin{eqnarray}}
\newcommand{\eea}{\end{eqnarray}}

\def\circa#1{\,\raise.3ex\hbox{$#1$\kern-.75em\lower1ex\hbox{$\sim$}}\,}

\DeclareSymbolFont{usualmathcal}{OMS}{cmsy}{m}{n}
\DeclareSymbolFontAlphabet{\mathcal}{usualmathcal}

\begin{document}

\begin{center}{\Large \textbf{
The $SU(2)_D$ lepton portals for muon $g-2$, $W$ boson mass and dark matter\\
}}\end{center}

\begin{center}
Seong-Sik Kim\textsuperscript{1},
Hyun Min Lee\textsuperscript{1},
Adriana G. Menkara\textsuperscript{1*} and
Kimiko Yamashita \textsuperscript{1}
\end{center}

\begin{center}
{\bf 1} Department of Physics, Chung-Ang University, Seoul 06974, Korea.
\\
* amenkara@cau.ac.kr
\end{center}

\begin{center}
\today
\end{center}


\definecolor{palegray}{gray}{0.95}
\begin{center}
\colorbox{palegray}{
  \begin{tabular}{rr}
  \begin{minipage}{0.1\textwidth}
    \includegraphics[width=30mm]{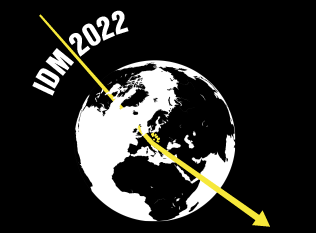}
  \end{minipage}
  &
  \begin{minipage}{0.85\textwidth}
    \begin{center}
    {\it 14th International Conference on Identification of Dark Matter}\\
    {\it Vienna, Austria, 18-22 July 2022} \\
    \end{center}
  \end{minipage}
\end{tabular}
}
\end{center}

\section*{Abstract}
{\bf

We propose a novel model which extends the Standard Model (SM) by introducing a $SU(2)_D$ gauge symmetry. In this model, a dark $SU(2)_D$ Higgs doublet and a Higgs bi-doublet can contribute to the muon g-2 anomaly and the $W$ boson mass, remaining in tune with the recent experimental results. At the same time, the isospin charged gauge boson of $SU(2)_D$ becomes a plausible candidate for Dark Matter(DM). We find that the resulting parameter space can fit the muon g-2, the W boson mass and the DM constraints simultaneously.
}


\vspace{10pt}
\noindent\rule{\textwidth}{1pt}
\tableofcontents\thispagestyle{fancy}
\noindent\rule{\textwidth}{1pt}
\vspace{10pt}


\section{Introduction}
\label{sec:intro}

Despite its success, the Standard Model (SM) still leaves us with some unanswered questions. One example is the well-stablished muon $g - 2$ anomaly from Brookhaven E821 \cite{Muong-2:2006rrc}, which has been recently confirmed by E989 \cite{Muong-2:2021ojo}. Another one is the latest measurement of the $W$ boson mass at Tevatron CDFII \cite{CDF:2022hxs}, which deviates with high significance from the SM prediction \cite{Haller:2018nnx}. On top of this, we have puzzling questions regarding Dark Matter(DM).  Although we have a large amount of indirect evidence for DM, such as galaxy rotation curves and the CMB, we still don't understand its nature neither its origin. The only hint we have is its total energy density, $\Omega h^2 \simeq 0 .1186$, as measured by Planck \cite{planck}. A connection between these puzzles could shed some light into the future experimental and theoretical steps to take.


\section{The model}
\label{sec:model}
In this work, we extend the SM by adding a new local $SU(2)_D$ symmetry. We introduced an $SU(2)_D$ double vector-like lepton $\Psi$, an $SU(2)_D$ doublet Higgs $\phi_D$ and a Higgs bi-doublet $H'$. The particle content of the model and their charges under the $Z_2$ parity are described in detail in reference \cite{Kim:2022zhj}.

The Lagrangian of our theory can be contains a DM and a Yukawa part including the vector-like leptons,

\bea
{\cal L}_{\rm DM} &=&-\frac{1}{2} {\rm Tr} \Big(V_{\mu\nu} V^{ \mu\nu}\Big)+i {\bar\Psi} \gamma^\mu D_\mu\Psi +|D_\mu\Phi_D|^2+{\rm Tr}\Big(|D_\mu H'|^2\Big) - V(\Phi_D,H',H) 
\eea

\bea
{\cal L}_{\rm VLSM}&=&-y_{d} {\bar q}_L d_R  H- y_{u} {\bar q}_L u_R {\tilde H}-y_{l} {\bar l}_L e_R H-y_{\nu} {\bar l}_L \nu_R {\tilde H}-M_R \overline{\nu^c_R}\nu_R   \nonumber \\
&&-M_{E} {\bar \Psi}\Psi-\lambda_{E}  {\bar \Psi}_L\Phi_D  e_{R}-y_{E}  {\bar l}_{L} H' \Psi_{R}  +{\rm h.c.}, \label{leptonL}
\eea

with the scalar potential given by
\bea
V(\Phi_D,H, H') &=& \mu^2_1 H^\dagger H + \mu^2_2 {\rm Tr}(H'^\dagger H')-( \mu_3 H^\dagger H' \Phi_D+{\rm h.c.})\nonumber \\
&+& \lambda_1 (H^\dagger H)^2 + \lambda_2 ({\rm Tr}H'^\dagger H')^2+ \lambda_3 (H^\dagger H){\rm Tr}(H'^\dagger H') \nonumber \\
&+& \mu^2_\phi \Phi^\dagger_D\Phi_D + \lambda_\phi (\Phi^\dagger_D\Phi_D)^2+ \lambda_{H\Phi}H^\dagger H\Phi^\dagger_D\Phi_D +  \lambda_{H'\Phi} {\rm Tr}(H'^\dagger H')\Phi^\dagger_D\Phi_D. \label{scalarpot}
\eea

Finally, the mass Lagrangian for the lepton sector is
\bea
{\cal L}_{L,{\rm mass}}= -M_E {\bar E}E-M_E {\bar E}' E' -m_0 {\bar e}e-( m_R {\bar E}_L e_R+m_L {\bar e}_L E_R+ {\rm h.c.}).
 \label{leptonmass0}
\eea

Diagonalizing the Lagrangian in equation (\ref{leptonmass0}), the vector-like lepton contributions are found to be naturally small \cite{kimiko,seesaw}, consistently with a simultaneous symmetry breaking of both the electroweak and the $SU(2)_D$ symmetries.

\section{Muon $g-2$ and  $W$  mass constraints}
\label{sec:muonandw}
In this section, we briefly summarize the contributions from the new particles in our model to the muon $g-2$ and the $W$ mass. 

\subsection{Muon $g-2$}

The dominant contributions to the muon $g-2$ are the ones coming from vector-like leptons and gauge bosons running together with muons in the loops. The vector contributions are given by \cite{seesaw,kimiko}
\bea
\Delta a^{V,E}_{\mu} \simeq  \left\{\begin{array}{c} \frac{g^2_D M_{E} m_{\mu}}{16 \pi^2 m_{V^0}^2}\, (c_{V}^2-c_{A}^2)+ \frac{g^2_D M_{E} m_{\mu}}{32 \pi^2 m_{V^0}^2}\, ({\hat c}_{V}^2-{\hat c}_{A}^2), \qquad M_{E}\gg m_{V^0}, \vspace{0.3cm} \\ \frac{g^2_D M_{E} m_{\mu}}{4 \pi^2 m_{V^0}^2}\, (c_{V}^2-c_{A}^2)+\frac{g^2_D M_{E} m_{\mu}}{8 \pi^2 m_{V^0}^2}\, ({\hat c}_{V}^2-{\hat c}_{A}^2), \quad m_{\mu}\ll M_{E}\ll m_{V^0}. \end{array}  \right. \label{zp1}
\eea
and 
\bea
\Delta a^{V^\pm,E'}_\mu =\frac{1}{2} \Delta a^{V^0,E}_l(c_V\to {\hat c}_V, c_A\to {\hat c}_A, M_E\to M_{E'}, m_{V^0}\to m_{V^\pm}),
\eea

while the contribution coming from the scalar $V_0$ reads \cite{seesaw,kimiko}
\bea
\Delta a^{V^0,\mu}_{\mu} &\simeq& \frac{g^2_D m^2_\mu}{12\pi^2 m^2_{V^0}}\, ( v^{\prime 2}_{\mu} -5 a^{\prime 2}_{\mu}).
\eea

Here, the coefficients $c_V, c_A, {\hat c}_V, {\hat c}_A,  v^{\prime }_{\mu} $ and $ a^{\prime }_{\mu} $  are given in eqs. (47)-(52) in \cite{Kim:2022zhj}.

\subsection{$W$ mass}
The dominant contribution for the $W$ mass is the tree-level process arising from the mixing between the gauge neutral bosons. The contribution to the $\rho$ parameter is then

\bea
\Delta \rho_H\simeq  \left\{ \begin{array}{cc} \frac{s^2_W g^2_D}{g^2_Y} \frac{M^2_Z}{m^2_{V^0}}\,\sin^4\beta, \quad  m_{V^0}\gg M_{Z}, \vspace{0.3cm} \\  -\frac{s^2_W g^2_D}{g^2_Y} \,\sin^4\beta, \quad m_{V^0}\ll M_Z. \end{array} \right.
\eea

We note that one-loop contributions from $E'$, ${\tilde\varphi}$ and $V^\pm$ are supressed by the small lepton mixing angle $\sin^2\zeta$. Consequently, they are sub-dominant compared to the tree-level effects of the $Z-V^0$ mass mixing and we can neglect them in our analysis.

\begin{figure}[t!]
\centering
\includegraphics[width=0.43\textwidth,clip]{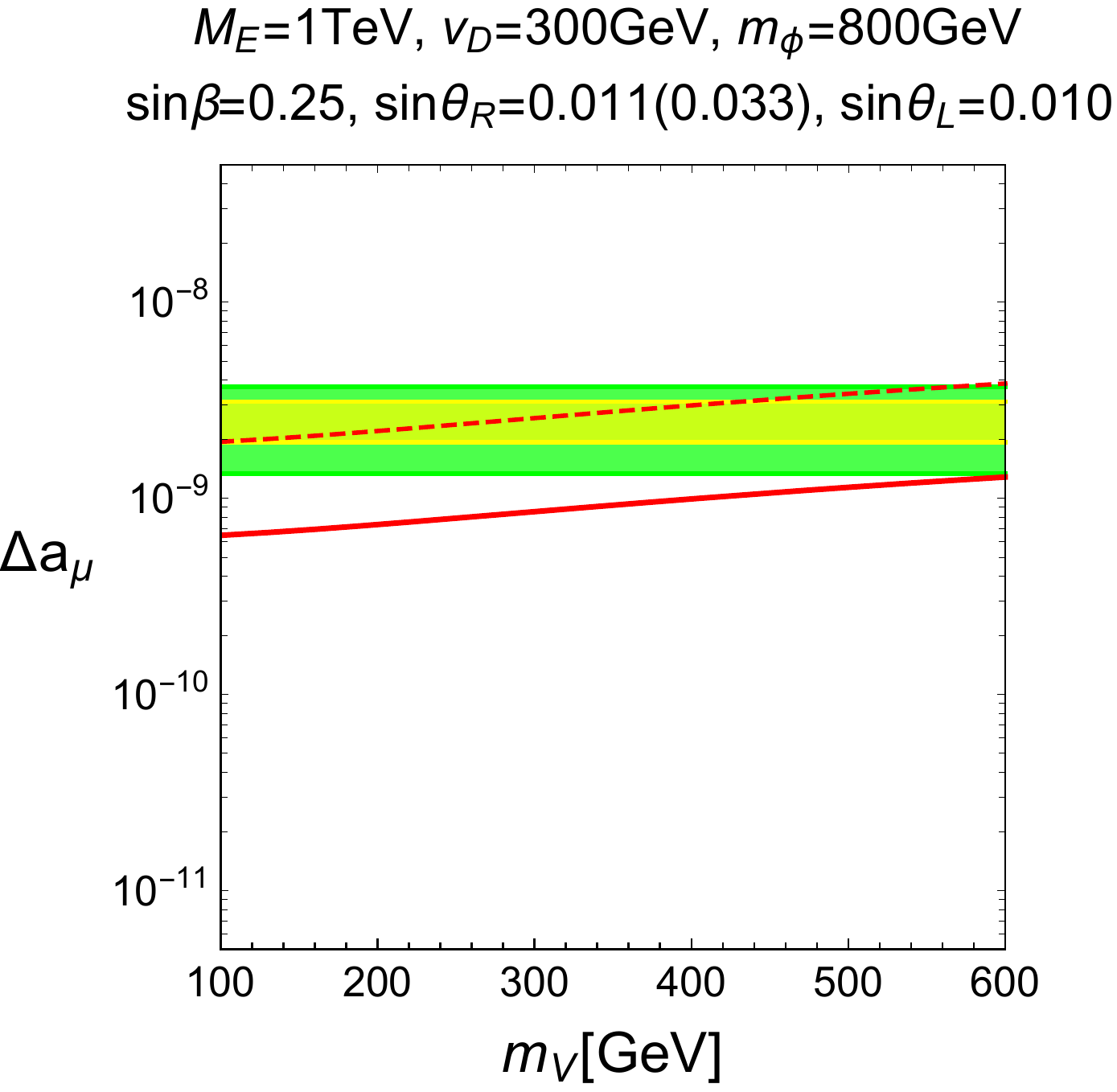}\,\,\,\,\,\,
\includegraphics[width=0.43\textwidth,clip]{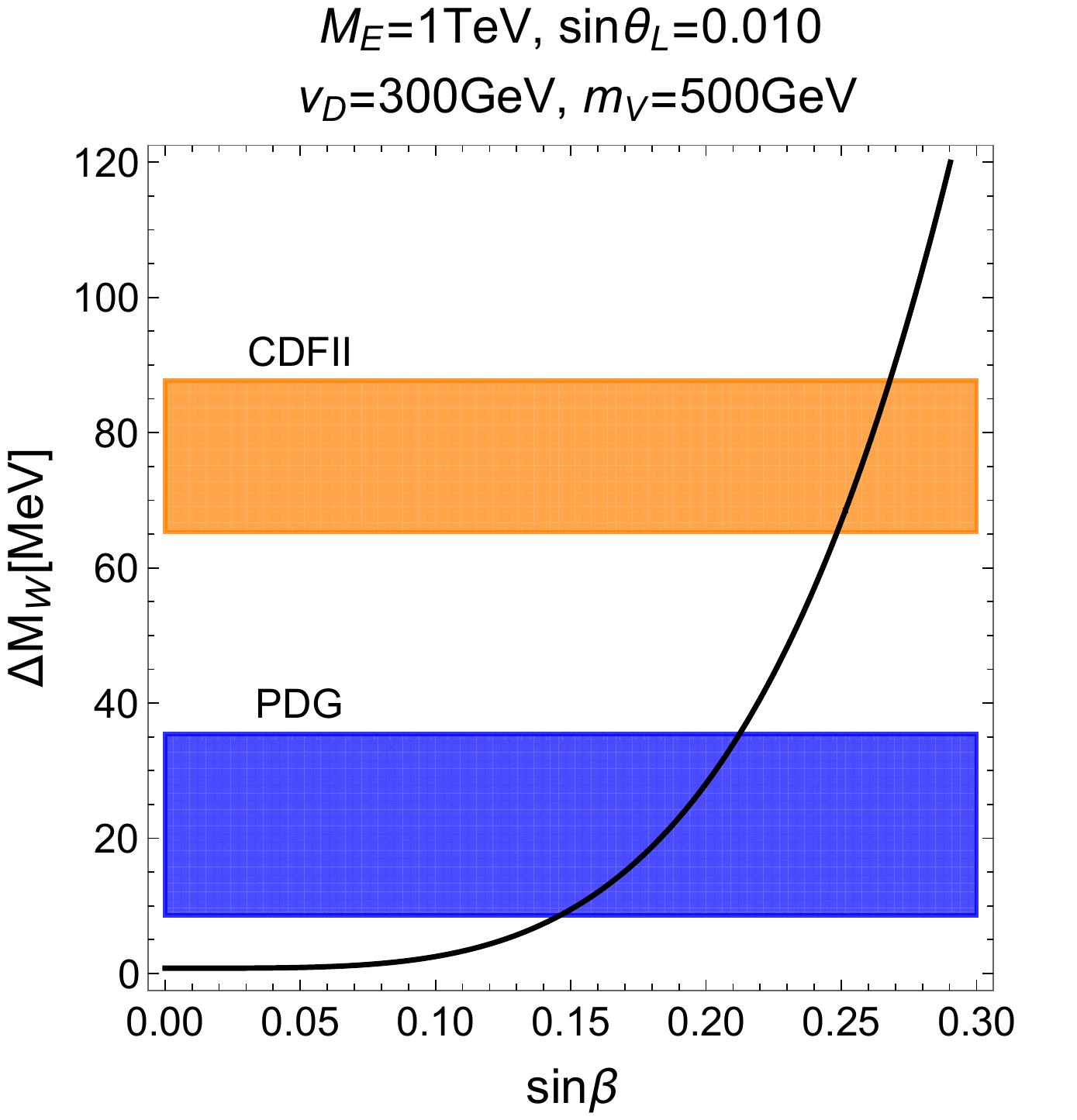}
\caption{Parameter space for the muon $g-2$ and the $W$ mass.}
\label{fig:mw}
\end{figure}

We summarize our results in Fig.~\ref{fig:mw}. On the left, we depict the muon $g-2$ correction as a function of $m_{V^+}$. The regions favored by the muon $g-2$ within $1\sigma$($2\sigma$) are shown in yellow(green). We find it necessary to set $m_\phi$ around the TeV scale in order to counter the ${\tilde\varphi}$ negative chirality enhanced contribution \cite{Kim:2022zhj}. On the right, we show the correction to the $W$ boson mass as a function of $\sin\beta$.

\section{Dark matter constraints}
\label{sec:darkmatter}

In this section we study whether our model can reproduce the correct DM relic abundance and satisfy the stringent bounds set by direct detection experiments. 

\subsection{Relic Density}
\label{sec:dmrelic}

The gauge bosons $V^{\pm}$, the vector-like lepton $E'$ and the neutral scalar $\tilde{\varphi}$ are odd under the $Z_2$ parity, and thus make viable DM candidates. In paticular, the $SU(2)_D$ gauge bosons are almost degenerate, but their masses receive a small positive correction proportional to the $Z - V^0$ mixing\footnote{In this sense, the mass splitting, $\delta \equiv m_{V^0}/m_{V^+} -1$, is closely related to the W boson mass.}. This results in a forbidden channel, $V^+ V^- \to V^0 V^0$, which is responsible for the correct relic energy density.

On the left plot of Fig.~\ref{fig:relic}, we show the relic abundance for DM as a function $\delta\equiv m_{V^0}/m_{V^+}-1$. The choice of parameters is consistent with both the muon $g-2$ and the $W$ boson mass results in the previous section. We find that the condition for a correct relic density \cite{planck} depends crucially on the mass splitting $\delta$ and the DM mass $m_{V^+}$ for a given $v_D$. 

\subsection{Indirect Detection}
We remark that the forbidden channel is closed for small DM velocities, namely $v_{\rm rel}\lesssim \sqrt{8\delta}$.  The velocity of DM in our galaxy, $v_{\rm rel}\simeq 220\,{\rm km}$  makes any signal from this channel unobservable for  $\delta\gtrsim 6\times 10^{-7}$. However, other subdominant channels, such as $V^+ V^-\to hh, V^0 Z$ or $V^+V^-\to {\rm SM}\,{\rm SM}$, could lead to interesting signals for indirect detection, for example in the CMB or in cosmic rays \cite{fmdm,srdm}. 


\subsection{Direct detection}

Decays of DM into a quark-antiquark pair, $V^+V^-\to q{\bar q}$,  are subdominant for the relic density abundance, but they can be constrained by  direct detection experiments. We find that the XENON1T bound \cite{xenon1t} on the DM-nucleon scattering cross section can be satisfied in the alignment limit for the Higgs mixing angle,  $\sin\theta_h\simeq -\frac{v}{\sqrt{2}v_D} \,\sin^2\beta$ and $m_s\gg m_h$.

\begin{figure}[t]
\centering
\includegraphics[width=0.33\textwidth,clip]{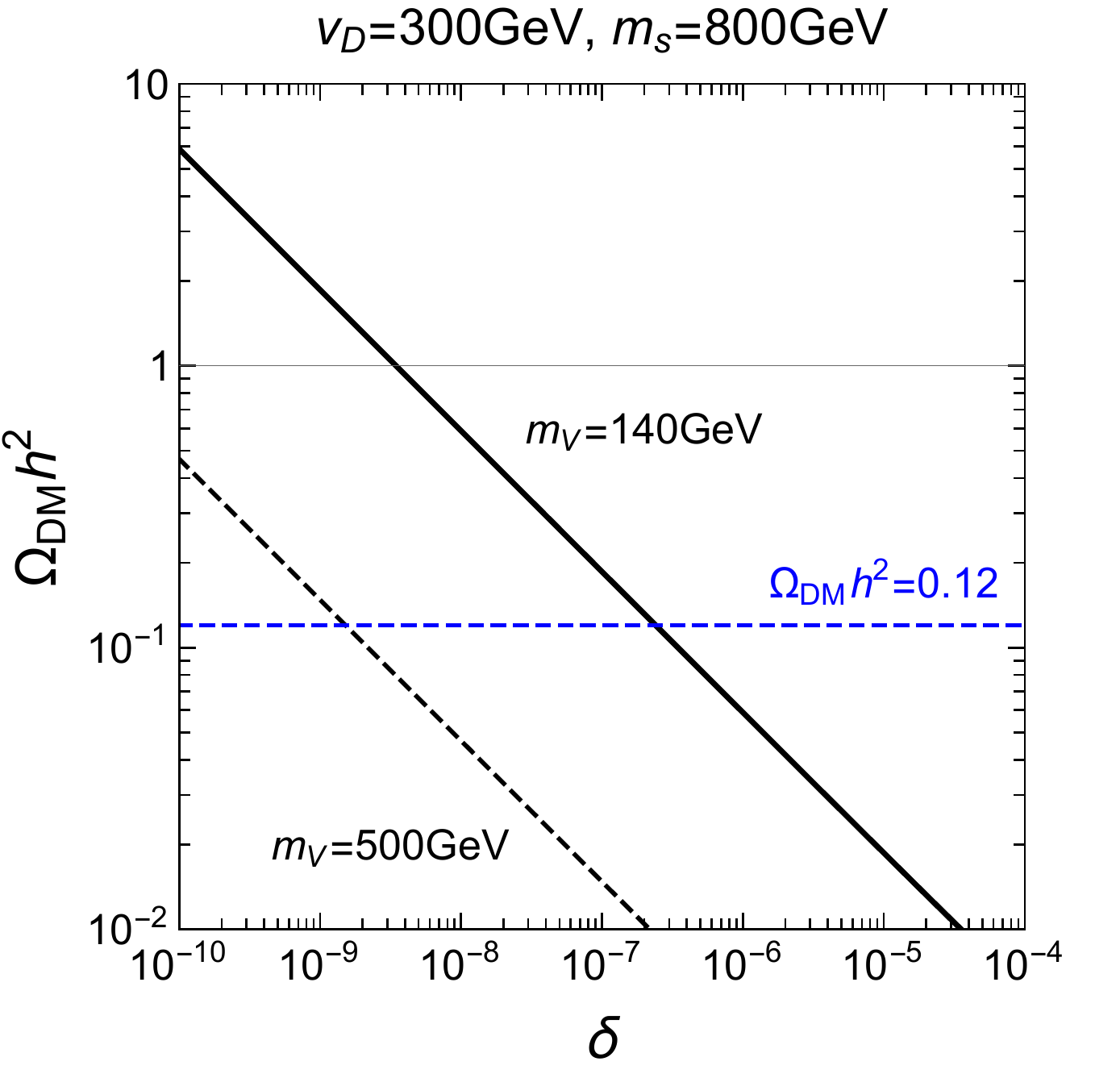}\,\,\,\,\,\,
\includegraphics[width=0.33\textwidth,clip]{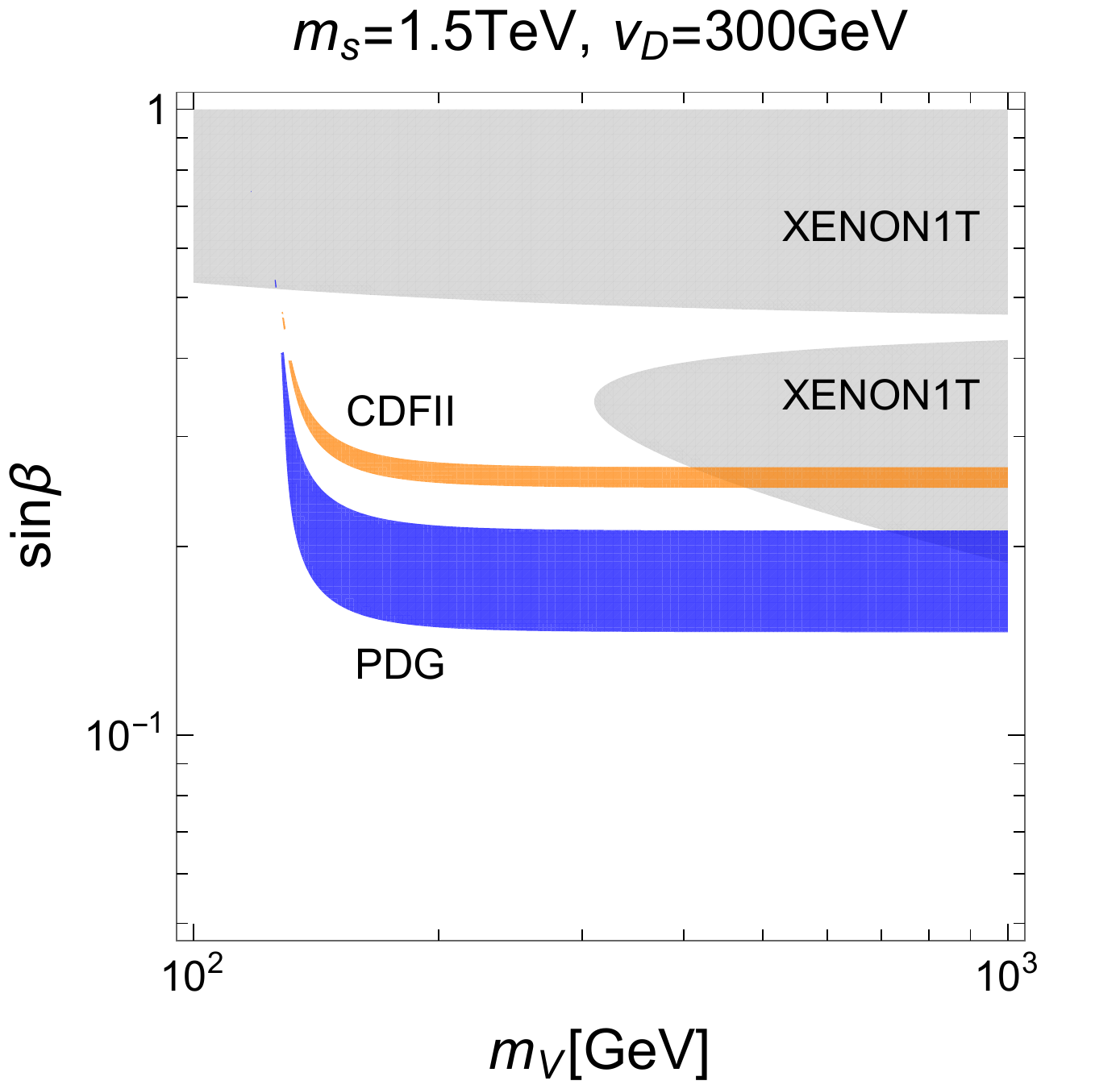}
\caption{(Left) DM relic density as a function of $\delta\equiv m_{V^0}/m_{V^+}-1$. (Right) Parameter space consistent with XENON1T\cite{xenon1t}.
}
\label{fig:relic}
\end{figure}

On the right plot of Fig.~\ref{fig:relic}, we show in gray the parameter space which has been ruled out by XENON1T\cite{xenon1t}.  In the alignment limit and for a heavy singlet scalar with $m_s=1.5\,{\rm TeV}$, our results show that there is consistent parameter space satisfying the muon $g-2$, the $W$ boson mass and the DM constraints simultaneously.

\section{Conclusions}

In this paper, we addressed the possibility of explaining DM and flavor puzzles simultaneously. With that in mind, we proposed a model that extends the SM by implementing a new $SU(2)_D$ gauge symmetry. The flavor mixing between the lepton and the vector-like lepton makes the lepton masses naturally small. Furthermore, the vector-like leptons and $SU(2)_D$ gauge bosons interactions contribute dominantly to the muon $g-2$, while the $Z-V^0$ mass mixing accounts for the deviation of the $W$ boson mass, as recently measured by CDFII. A combination of the $U(1)_G$ global symmetry in the Higgs sector and the dark isospin symmetry leads to a $Z_2$ parity, allowing for suitable DM candidates. We found that the correct DM relic density can be reproduced at the same time that direct detection bounds are satisfied. Within the same parameter space we can also explain the muon $g-2$ and the $W$ boson mass anomalies.

\paragraph{Funding information}
This work is partly supported by  the Basic Science Research Program through the National
Research Foundation of Korea (NRF) funded by the Ministry of Education, Science and
Technology (NRF-2022R1A2C2003567 and NRF-2021R1A4A2001897). 
The work of KY is supported by Brain Pool program funded by the Ministry of Science and ICT through the National Research Foundation of Korea(NRF-2021H1D3A2A02038697).


\nolinenumbers

\end{document}